\documentclass[preprint,showpacs,preprintnumbers,amsmath,amssymb]{revtex4}

\usepackage{graphicx}
\usepackage{dcolumn}
\usepackage{bm}

\begin{document}

\title{Stochastic dynamics and mean field approach in a system of three interacting species}
\author{\large Davide Valenti\footnote{e-mail: valentid@gip.dft.unipa.it} and Bernardo Spagnolo\footnote{e-mail: spagnolo@unipa.it}}
\affiliation{Dipartimento di Fisica e Tecnologie Relative, Group of
Interdisciplinary Physics\footnote{Electronic address:
http://gip.dft.unipa.it}, Universit\`a di Palermo and CNISM-INFM\\
Viale delle Scienze, I-90128 Palermo, Italy}

\begin{abstract}
The spatio-temporal dynamics of three interacting species, two
preys and one predator, in the presence of two different kinds of
noise sources is studied. To describe the spatial distributions of
the species we use a model based on Lotka-Volterra equations. A
correlated dichotomous noise acts on $\beta$, the interaction
parameter between the two preys, and a multiplicative white noise
affects directly the dynamics of each one of the three species. We
study the time behaviour of the three species in single site for
different values of the multiplicative noise intensity, finding
noise-induced oscillations of the three species densities with an
anticorrelated behaviour of the two preys. Afterwards, by
considering a spatially extended system formed by a two-dimensional
lattice with $N$ sites and applying a mean field approach, we get
the corresponding moment equations in Gaussian approximation. Within
this formalism we obtain the time behaviour of the first and second
order moments for different values of multiplicative noise
intensity, with $\beta(t)$ subject to the same dichotomous noise
source. Finally, we compare our results with those obtained by using
a coupled map lattice model, consisting of a time discrete version
of the Lotka-Volterra equations.
\end{abstract}


\maketitle


\section{Introduction}
\label{intro} \vskip-0.2cm Noise is not generally detrimental to
biological systems but can be employed to generate genotypic,
phenotypic, and behavioral
diversity~\cite{Hoffmann,Korobkova,Thattai,Samoilov}. Real
ecosystems are affected by the presence of noise sources which
consist of random variability of environmental parameters, such as
temperature, food availability, general conditions which can favour
or thwart the increase of some biological species. This randomly
fluctuating behaviour can be modeled by Gaussian noise sources,
which influence, through a multiplicative interaction, the system
dynamics. Multiplicative noise often causes the appearance of
fluctuating barriers or processes of anomalous diffusion and has
been investigated in the context of population growth and
extinction~\cite{Hoffmann,Zimmer,Ciuchi,Biro,Fa,Kaniadakis,Hanggi,Fleming,Vladar,Wichmann,Ai,Halley}.
In this paper we study the time evolution of three interacting
species, two preys, $x$ and $y$, and one predator, $z$. The
interaction between the two preys is symmetric and it is given by
the parameter $\beta$. We study the ecosystem dynamics, described by
generalized Lotka-Volterra equations, in the presence of two
different kinds of noise sources: (i) a dichotomous noise acting on
the $\beta$ parameter, (ii) three external sources, modeled as
independent multiplicative Gaussian noises, which act directly on
the three species. First we consider the deterministic dynamics of
the system in a single site and we get the time behaviour of $x$,
$y$ and $z$, by analyzing the stability of the ecosystem with
different constant values of the interaction parameter $\beta$,
which correspond to a coexistence regime ($\beta_{down}<1$) or to an
exclusion regime ($\beta_{up}>1$). Then we consider the interaction
parameter $\beta$ varying dichotomously between these two values. In
this condition we study the time behaviour of the species
concentrations $x$, $y$ and $z$ for different levels of the
multiplicative noise intensity. We find noise-induced oscillations
and strong anticorrelations between the preys. Afterwards we take
into account the spatial version of our ecosystem, considering a
two-dimensional domain formed by $N$ sites and adding a diffusion
term in the L-V equations. By using a mean field approach, we obtain
the corresponding moment equations in Gaussian approximation. We
find that, for $\beta$ varying dichotomously, the $1^{st}$ order
moments of the three species concentrations are independent on the
multiplicative noise intensity. On the other hand, the behavior of
the $2^{nd}$ order moments is strongly affected by the presence of
external noise sources. In particular we find that the time behavior
is anticorrelated for the species densities of the two preys, and
correlated between the predator and the total density of the two
preys. Finally we get the time behavior of the $1^{st}$ and $2^{nd}$
order moments using a coupled map lattice (CML) model~\cite{Kaneko}
and we compare these results with those previously obtained within
the mean field approach. In view of an application on real systems,
the results obtained could be useful to explain experimental data,
reproducing the behaviour of natural
ecosystems~\cite{Zimmer,Ciuchi,Spagnolo,Garcia}.

\section{The model}
\label{sec:1} \vskip-0.2cm Our system is described by a time
evolution model of Lotka-Volterra equations, within the Ito scheme,
with diffusive terms in a spatial lattice consisting of $N$ sites
\begin{eqnarray}
\dot{x}_{i,j}&=&\lambda \thinspace x_{i,j}
\thinspace(1-\nu\thinspace x_{i,j}-\beta\thinspace
y_{i,j}-\alpha\thinspace z_{i,j})+x_{i,j} \thinspace
\sqrt{\sigma_x}\thinspace\xi^x_{i,j} + D(<x>-x_{i,j})
\label{Lotka_eq_1}\\
\dot{y}_{i,j}&=&\lambda \thinspace y_{i,j} \thinspace(1-\nu
\thinspace y_{i,j}-\beta \thinspace x_{i,j} -\alpha \thinspace
z_{i,j})+ y_{i,j} \thinspace \sqrt{\sigma_y} \thinspace\xi^y_{i,j} +
D(<y>-y_{i,j})
\label{Lotka_eq_2}\\
\dot{z}_{i,j}&=&\lambda_z \thinspace z_{i,j}\thinspace
[-1+\gamma\thinspace(x_{i,j}+y_{i,j})] +
z_{i,j}\thinspace\sqrt{\sigma_z}\thinspace\xi^z_{i,j} +
D(<z>-z_{i,j}),\label{Lotka_eq_3}
\end{eqnarray}

where the dot indicates the time derivative. The variables
$x_{i,j}$, $y_{i,j}$ and $z_{i,j}$ are functions of the time $t$,
and denote the densities, respectively, of the two preys and the
predator in the lattice site $(i,j)$. $\lambda$ and $\lambda_z$ are
scale factors, $\nu$ is the growth rate for the two preys, $D$ is
the diffusion coefficient, and $<x>$, $<y>$, $<z>$ indicate the
spatial mean, performed on the whole lattice, of the three species
densities. The coefficient $\beta$ is the interaction parameter
between the two preys. The coefficients $\alpha$ and $\gamma$
account for the interaction between preys and predator.
$\xi^x_{i,j}(t)$, $\xi^y_{i,j}(t)$, $\xi_{i,j}^z(t)$ are
statistically independent Gaussian white noises with zero mean and
unit variance, and they model the interaction between species and
environment. Finally, $\sigma_x$, $\sigma_y$, $\sigma_z$ are the
intensities of the three sources of Gaussian white noise.

\subsection{Single site dynamics}
\label{single site dynamics}
\subsubsection{Stability analysis and dynamical regimes}
\vskip-0.2cm

\begin{figure}
\begin{center}
\includegraphics[width=13cm]{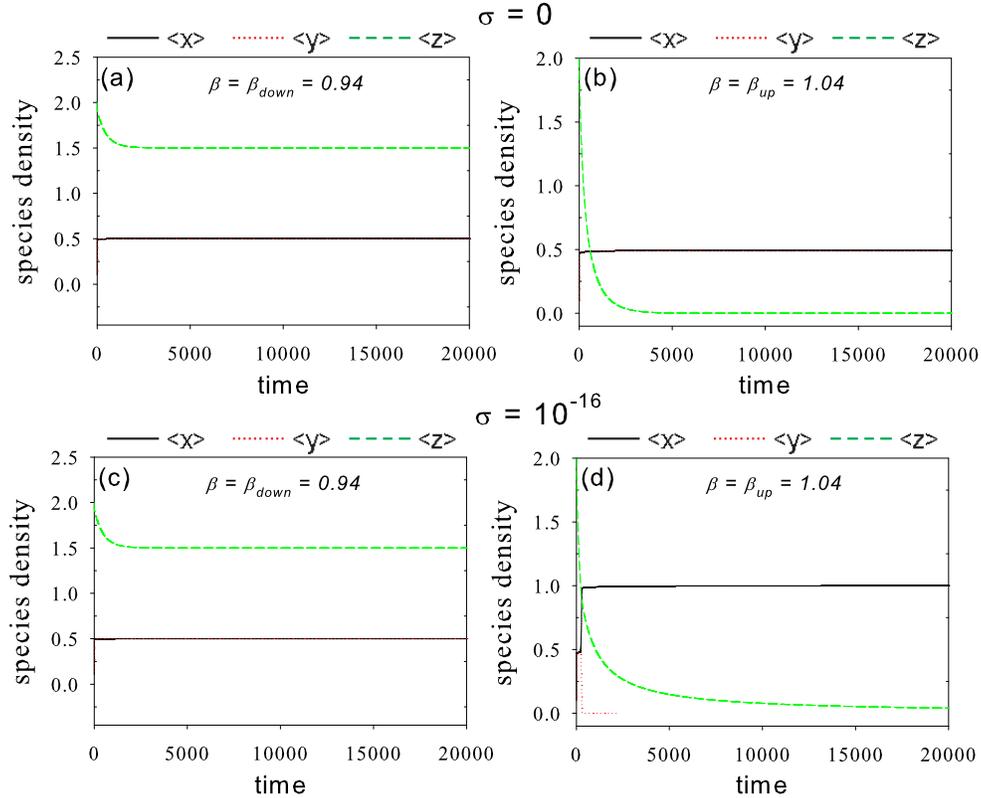}
\end{center}
\caption{Time evolution of the three species. Deterministic dynamics
in (a) coexistence and (b) exclusion regime. Stochastic dynamics,
for $\sigma=10^{-16}$ in (c) coexistence and (d) exclusion regime.
Values of the parameters and initial conditions are $\lambda=3$,
$\lambda_z=0.06$, $\nu=1$, $\alpha=0.02$, $\gamma=1$, $x(0)$ =
$y(0)$ = $0.1$, $z(0)$ = $2.0$.} \label{beta=const} \vskip-0.3cm
\end{figure}

Depending on the value of the interaction parameter, coexistence or
exclusion regimes take place. In the absence both of multiplicative
noise ($\sigma_x=\sigma_y=\sigma_z=0$) and diffusion terms ($D=0$),
Eqs.~(\ref{Lotka_eq_1})-(\ref{Lotka_eq_3}) describe the
deterministic dynamics of a single site ecosystem. In these
conditions, for the generic site of lattice the stationary values of
the three species densities are given by
\begin{eqnarray}
x^{stat}&=& y^{stat} = \frac{1}{2\gamma} \label{x_stat}\\
z^{stat}&=&\frac{2\gamma-(\beta+\nu)}{2\alpha\gamma}.\label{z_stat}
\end{eqnarray}
where the indices $i,j$ where suppressed. From
Eqs.~(\ref{Lotka_eq_1})-(\ref{Lotka_eq_3}) one can see that the two
prey densities have stationary values that are independent on the
interaction parameter $\beta$. Conversely, the stationary value of
the predator density is connected with the value of $\beta$. This
indicates that the interaction parameter between the two preys
determines the coexistence or exclusion regimes for the whole
system, affecting the stationary value $z^{stat}$. From
Eq.~(\ref{z_stat}) the survivance condition for the predator is
$z^{stat} > 0$, which allows to get the coexistence condition for
the three species as a function of $\beta$
\begin{equation}
\beta<2\gamma-\nu.
\label{coex_2}
\end{equation}
The inequality (\ref{coex_2}) indicates that the system is
characterized by two stationary states, which become stable or
unstable depending on the values that $\beta$, $\gamma$ and $\nu$
take on. In particular, when the condition~(\ref{coex_2}) is
satisfied, the stable state is represented by the coexistence of the
three species. Otherwise, after a transient, the predator tends to
disappear (inequality (\ref{coex_2}) doesn't hold anymore) and we
get a system formed by two competing species, whose
coexistence/exclusion conditions depend directly on the value of the
parameter
$\beta$~\cite{ValSch,Spagnolo,Vilar-Spagnolo,Valenti,Valenti1}. In
this sense, the predator plays a regulatory role for the dynamics of
the two preys, whose reciprocal behavior is mediated by the
interaction parameter $\beta$ through the presence of the species
$z$. We calculate the numerical solutions for single site dynamics,
setting in Eqs.~(\ref{Lotka_eq_1})~-~(\ref{Lotka_eq_3}) $\lambda=3$,
$\lambda_z=0.06$, $\nu=1$, $\alpha=0.02$, $\gamma=1$, with two
different values $\beta_{down}=0.94$ and $\beta_{up}=1.04$ of the
interaction parameter $\beta$ and initial conditions $x(0)$ = $y(0)$
= $0.1$, $z(0)$ = $2.0$. The values of multiplicative noise
intensity are the same for the three species, that is
$\sigma=\sigma_x=\sigma_y=\sigma_z$. In Fig.~\ref{beta=const} we
show the time series of the three species in coexistence
($\beta=\beta_{down}$) and exclusion ($\beta=\beta_{up}$) regimes,
for $\sigma=0$ and $\sigma=10^{-16}$. We note that, when the system
is subject to deterministic dynamics, the coexistence regime causes,
after a transient, the three species to reach the equilibrium
values, $x^{stat}$ = $y^{stat}$ = $0.5$, $z^{stat}$ = $1.5$,
obtained from Eqs.~(\ref{x_stat})~-~(\ref{z_stat}) using $\nu=1$,
$\alpha=0.02$, $\gamma=1$, $\beta=\beta_{down}=0.94$ (see
Fig.~\ref{beta=const}a). In deterministic exclusion regime the
predator tends very slowly to vanish. However, the two prey
densities reach the stationary values, remaining constant
(Fig.~\ref{beta=const}b). In this case, the stationary values
correspond to an unstable equilibrium point. In fact, in the
presence of a small level of multiplicative noise, the symmetry, due
to the parameter values and initial conditions used in our
simulations, is broken and one of the two preys prevails, displacing
the other one (Fig.~\ref{beta=const}d), according to the previously
obtained results~\cite{Valenti}. Finally we note that no
significative modifications occur, with respect to the deterministic
case, when a small level of noise is present in coexistence regime
(see Fig.~\ref{beta=const}c). This obviously depends on the fact
that, for $\beta=\beta_{down}$, the system occupies a stable
equilibrium point, which is maintained also in the presence of low
levels of multiplicative noise.
However, environmental perturbations, due to the
presence both of deterministic and random fluctuations of biological
and physical variables, such as the temperature, affect the dynamics
of the species. These external forces can modify the behaviour of
the populations, either introducing multiplicative noise sources
which act directly on the species or affecting the dynamics of the
interaction parameter $\beta$. In fact, the environmental variations
can cause the system dynamics to change between coexistence
($\beta<2\gamma-\nu$) and exclusion ($\beta>2\gamma-\nu$) regimes.
This dynamical behavior can be described by considering that the
interaction parameter $\beta(t)$ is a stochastic process driven by a
dichotomous noise, whose jump rate is given by
\begin{align}
\chi(t) = \left\{
\begin{array}
[c]{ll}%
0, &\quad  \Delta t \leq \tau_d\\
\chi_0 \left(1 + A \thinspace \vert \cos\omega t \vert \right),
&\quad \Delta t > \tau_d\;.
\end{array}
\right. \label{jump_rate}
\end{align}

where $\Delta t$ is the time interval between two consecutive
switches, and $\tau_d$ is the delay between two jumps, that is the
time interval after a switch, before another jump can occur.
\begin{figure}
\begin{center}
\includegraphics[width=6cm]{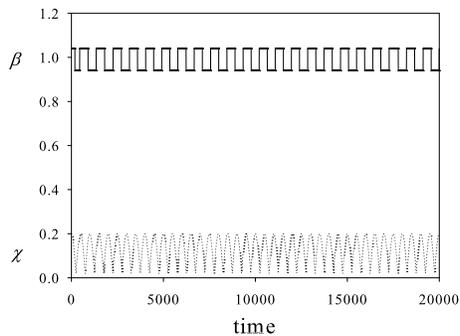}
\end{center}
\caption{Time evolution of the interaction parameter $\beta(t)$ with
initial value $\beta(0)=1.04$ and delay time $\tau_d = 435$. The
interaction parameter $\beta(t)$ switches quasi-periodically between
$\beta_{down}=0.94$ and $\beta_{up}=1.04$. The values of the other
parameters are $A =9.0$, $\omega/(2\pi)=10^{-3}$,
$\chi_0=2\cdot10^{-2}$.} \label{beta} \vskip-0.3cm
\end{figure}
In Eq.~(\ref{jump_rate}), $A$ and $\omega = (2\pi)/T$ are
respectively amplitude and angular frequency of the periodic term,
and $\chi_0$ is the jump rate in the absence of periodic term. This
causes $\beta(t)$ to jump between two values,
$\beta_{down}<2\gamma-\nu$ and $\beta_{up}>$~$2\gamma-\nu$.
According to the condition~(\ref{coex_2}), these values determine
the two possible dynamical regimes (coexistence or exclusion) of the
deterministic Lotka-Volterra's model for three interacting species.
For given values of the parameters $A$, $\omega$ and $\chi_0$ the
switching time between the two levels of $\beta(t)$ depends on
$\tau_d$. Applying a procedure analogous to that followed for the
two-species case~\cite{ValSch}, we set $A =9.0$,
$\omega/(2\pi)=10^{-3}$, $\chi_0=2\cdot10^{-2}$, obtaining the time
series of $\beta(t)$ for $\tau_d=435$, with $\beta_{down} = 0.94$
and $\beta_{up} = 1.04$. The results, shown in Fig.~\ref{beta},
indicate the presence of a synchronization between the jumps and the
periodicity of the rate $\chi(t)$. For a system formed by two
competing species this causes a quasi-periodical time behavior of
the two populations, which can be considered as a signature of the
stochastic resonance phenomenon~\cite{Benzi} in population
dynamics~\cite{Vilar-Spagnolo,Valenti,Valenti1}. Therefore we fix
the delay at the value $\tau_D=435$, which determines an oscillating
dynamical regime. In these conditions, $\beta(t)$ switches
quasi-periodically between $\beta_{down}$ and $\beta_{up}$ (see
Fig.~\ref{beta}), causing the system to be alternatively subject to
the coexistence and exclusion regimes.

\subsubsection{Time behaviour of the species in a single site}

\vskip-0.2cm In this section we analyze the time behaviour of three
interacting species in a single site of the lattice. From
Eqs.~(\ref{Lotka_eq_1})-(\ref{Lotka_eq_3}), by setting $D=0$ we get
\begin{eqnarray}
\dot{x}&=&\lambda \thinspace x \thinspace(1-\nu\thinspace
x-\beta\thinspace y-\alpha \thinspace z)+x
\thinspace\sqrt{\sigma_x}\thinspace\xi^x(t)\label{Lotka_eq_1_bis}\\
\dot{y}&=&\lambda \thinspace y \thinspace(1-\nu \thinspace y-\beta
\thinspace x -\alpha \thinspace z)+ y \thinspace
\sqrt{\sigma_y}\thinspace \xi^y(t)\label{Lotka_eq_2_bis}\\
\dot{z}&=&\lambda_z \thinspace z \thinspace (-1+\gamma \thinspace
x+\gamma \thinspace y) +z\thinspace\sqrt{\sigma_z}\thinspace
\xi^z(t),\label{Lotka_eq_3_bis}
\end{eqnarray}
where the indices $i,j$ where suppressed.
\begin{figure}[htbp]
\begin{center}
\includegraphics[width=13cm]{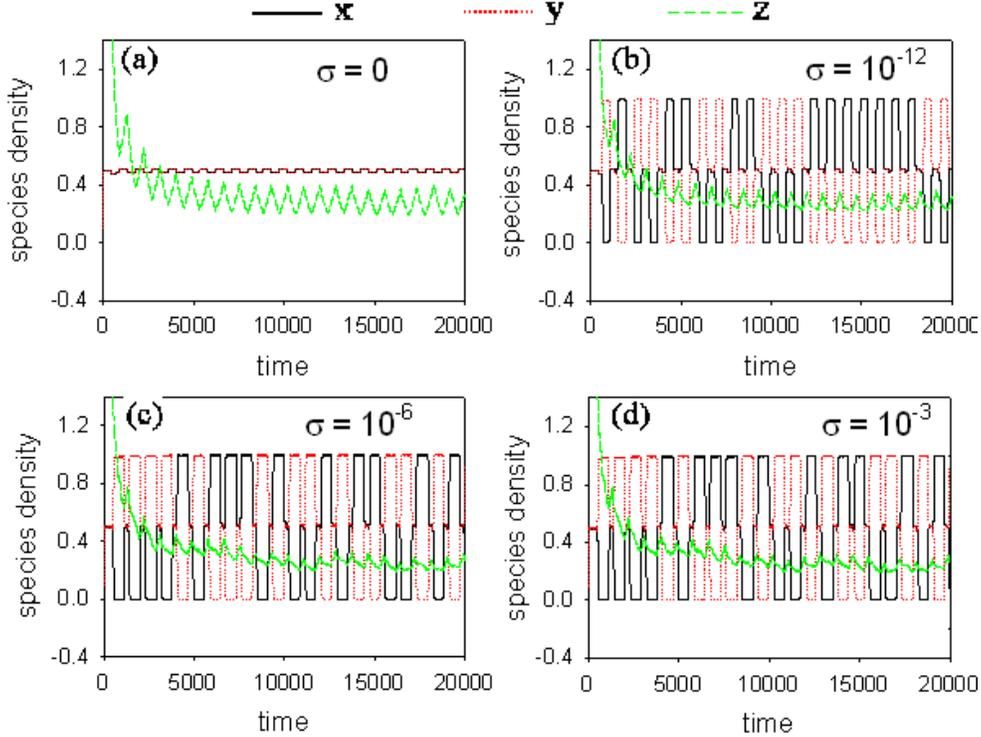}
\end{center}
\vskip-0.3cm \caption{\small Time evolution of the three species
densities in a single site of the lattice. The values of the
multiplicative noise intensity are: (a) $\sigma=0$, (b)
$\sigma=10^{-12}$, (c) $\sigma=10^{-6}$, (d) $\sigma=10^{-3}$. Here
$\lambda=3$, $\lambda_z=0.06$, $\nu=1$, $\alpha=0.02$, $\gamma=1$.
The values of the other parameters are the same of Fig.~\ref{beta}.
The initial values of the species densities are $x(0)$ = $y(0)$ =
$0.1$, $z(0)$ = $2.0$. The time series of $x(t)$, $y(t)$ (preys) and
$z(t)$ (predator) show a correlated behaviour in the absence of
noise (panel a). In the presence of the noise (panels (b)-(d)) an
anticorrelated behaviour of $x(t)$ and $y(t)$ appears.\bigskip}
 \label{single_compartment}
 \vskip-0.3cm
\end{figure}
%

By choosing $\beta(0) = 1.04$ and $\tau_d = 435$, we obtain for
$\beta(t)$ the time behaviour shown in Fig.~\ref{beta}. We analyze
the time evolution of the species densities by numerical simulation
of Eqs.~(\ref{Lotka_eq_1_bis})-(\ref{Lotka_eq_3_bis}). The time
series of $x$, $y$ and $z$ are obtained for different values of the
multiplicative noise intensity, namely $\sigma = 0$, $10^{-12}$,
$10^{-6}$, $10^{-3}$. The values of the other parameters are the
same used in the previous section, that is $\lambda=3$,
$\lambda_z=0.06$, $\nu=1$, $\alpha=0.02$, $\gamma = 1$,
$\beta_{down}=0.94$, $\beta_{up}=1.04$. The initial values of the
species densities are $x(0)$ = $y(0)$ = 0.1, $z(0)$ = $2.0$. In
Fig.~\ref{single_compartment}, where the results are reported, the
time series of $x(t)$, $y(t)$ (preys) and $z(t)$ (predator) show
correlated behaviour in the absence of noise (panel a). In the
presence of noise intensity an anticorrelated oscillating behaviour
of $x(t)$ and $y(t)$ appears (see panels (b)-(d)). Moreover we note
that, for all the values of multiplicative noise intensity, the two
prey densities oscillate, with the frequency of the external driving
force, around the stationary values, $x^{stat}=y^{stat}=0.5$. We
observe that the predator density show an oscillating behaviour,
with the same frequency, around a value smaller than $z^{stat}=1.5$.
However, the oscillations of $z(t)$ are characterized by a larger
amplitude with respect to $<x(t)>$ and $<y(t)>$. This behaviour is
connected with the different effect that the alternating regime
(exclusion/coexistence) produces on preys and predator. In fact, the
quasi-periodical behaviour of $\beta(t)$ affects directly the
dynamics of the predator (see Eq.~\ref{z_stat}), causing a decrease
of the mean value of $z$ during the exclusion regime. Conversely, in
coexistence regime the two preys maintain a constant value (see
Eq.~(\ref{x_stat})) going towards an anticorrelated regime for
$\beta(t)=\beta_{up}$. In this last condition the two preys are
subject to a pure competitive dynamics, recovering the behaviour
observed in a system of two competing species~\cite{Valenti}.

\subsection{Spatially extended system: Mean field approach}
\label{spatially_extended} 
\begin{figure}[htbp]
\begin{center}
\includegraphics[width=15cm]{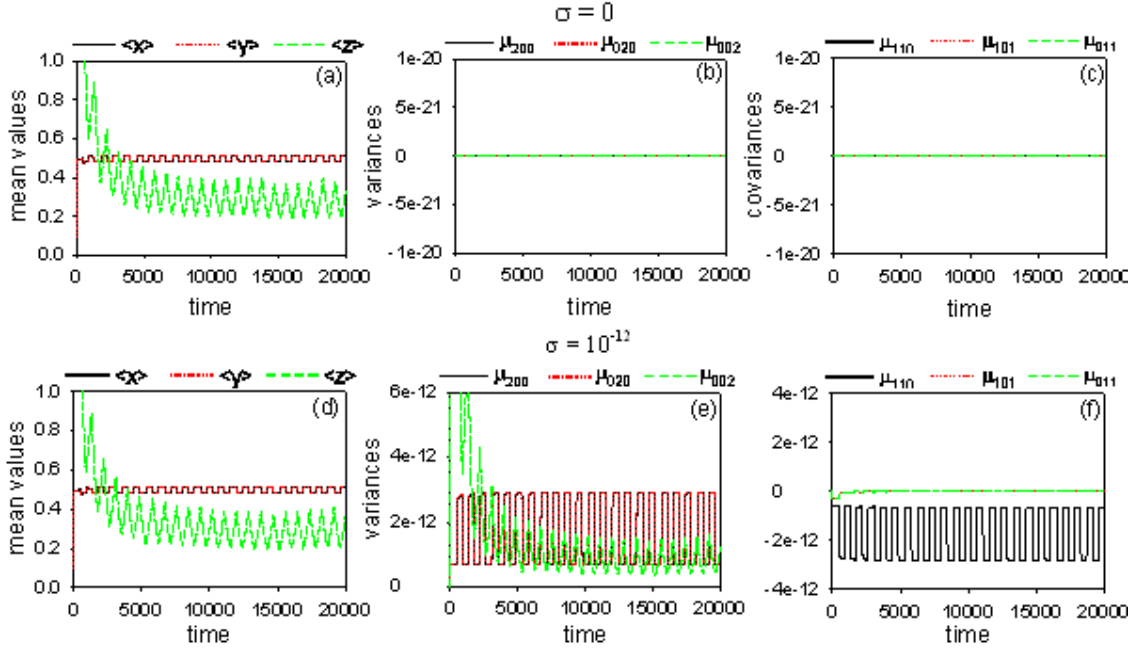}
\end{center}
\vskip-0.3cm \caption{\small Time evolution of the $1^{st}$ and
$2^{nd}$ order moments in the mean field approach. The values of the
multiplicative noise intensity are: $\sigma=0$, $10^{-12}$ from top
to bottom. In the absence of noise the time series of $<x(t)>$,
$<y(t)>$ (panel a), $\mu_{200}$, $\mu_{020}$, $\mu_{002}$ (panel b)
and $\mu_{110}$, $\mu_{101}$, $\mu_{011}$ (panel c) are completely
overlapped. The predator (mean value of species $z$) shows a
behaviour correlated with those of both preys (mean values of
species $x$ and $y$). For $\sigma=10^{-12}$, no changes are observed
in the behaviour of the mean values (panel d), the variances of the
two preys oscillate overlapping each other and a correlation is
observed with the variance of the species $z$ (panel e), the
covariance of the two preys, $\mu_{110}$, oscillates taking on only
negative values (the two preys are anticorrelated each other), while
$\mu_{101}$ and $\mu_{011}$ are always zero (panel f). The initial
values of the moments are $<x(0)>$ = $<y(0)>$ = $0.1$, $<z(0)>$ =
$2.0$, $\mu_{200}(0)$=$\mu_{020}(0)$=$\mu_{002}(0)$ =
$\mu_{110}(0)$=$\mu_{110}(0)=0$=$\mu_{011}(0)$=$0$. The diffusion
coefficient is $D=10^{-1}$. The values of the other parameters are
the same used in Fig.~\ref{single_compartment}.\bigskip}
\label{moment_series_1} \vskip-0.3cm
\end{figure}

\begin{figure}[htbp]
\begin{center}
\includegraphics[width=15cm]{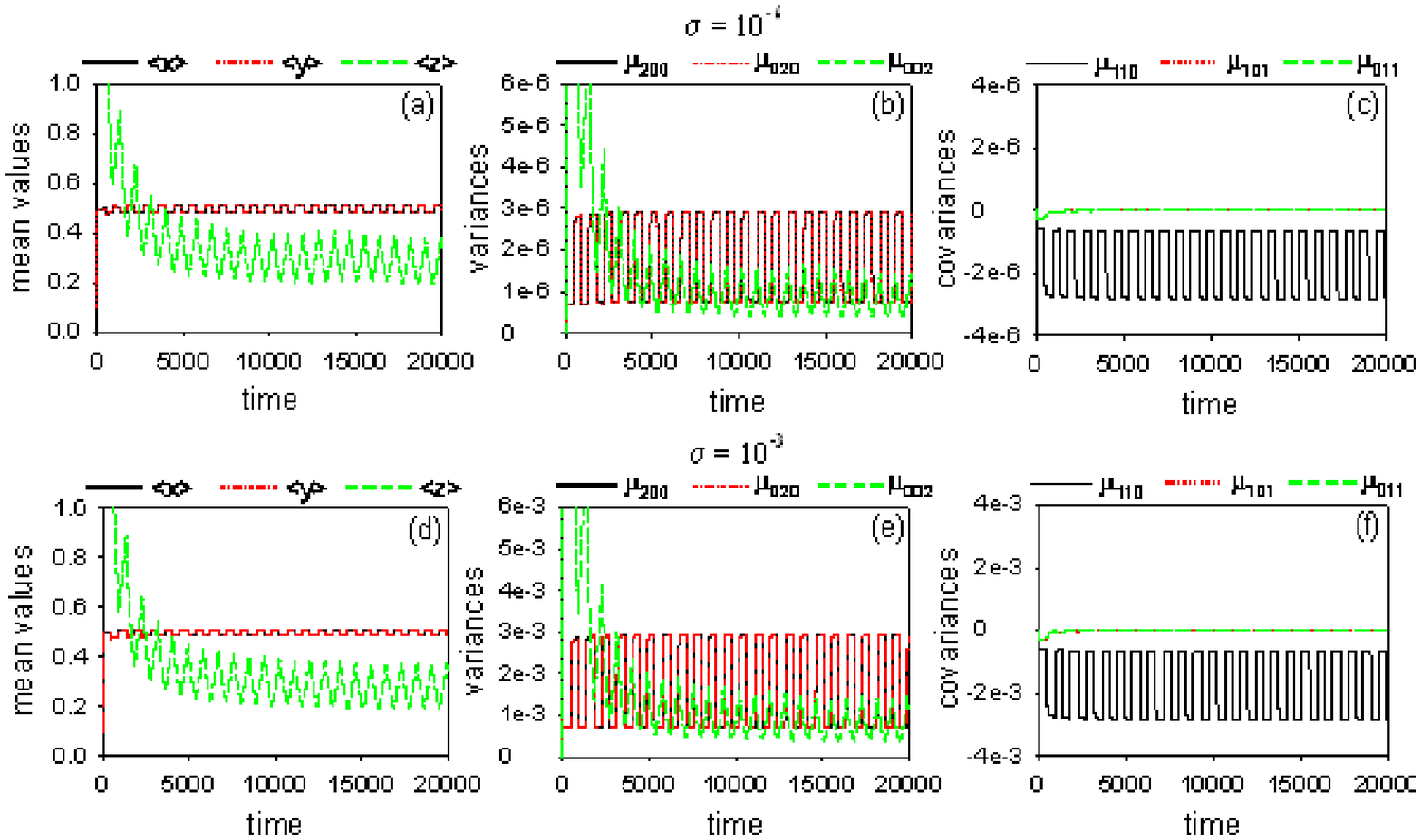}
\end{center}
\vskip-0.6cm \caption{\small Time evolution of the $1^{st}$ and
$2^{nd}$ order moments. The values of the multiplicative noise
intensity are: $10^{-6}$, $10^{-3}$ from top to bottom. No changes
are observed in the time behaviour of $<x>$ = $<y>$ = $<z>$ (see
panels a and d) for both values of the noise intensity. An increase
in the amplitude of oscillations, as a function of the noise
intensity, appears both in the variances of the two predators,
$\mu_{200}$, $\mu_{020}$, (see panels b and e) and in the covariance
of the two preys, $\mu_{110}$, (see panels c and f). The values both
of initial conditions and parameters are the same used in
Fig.~\ref{moment_series_2}.\bigskip} \label{moment_series_2}
\vskip-0.3cm
\end{figure}

\vskip-0.2cm In this section we analyze
the time behaviour of three interacting species in a spatially
extended system by using a mean field approach. The system dynamics
is described by Eqs.~(\ref{Lotka_eq_1})-(\ref{Lotka_eq_3}) in the
presence of the diffusive term ($D\neq0$). In order to use a mean
field approach we derive the moment equations for this system.
Assuming $N \rightarrow \infty$, we write
Eqs.~(\ref{Lotka_eq_1})-(\ref{Lotka_eq_3}) in a mean field form
\begin{eqnarray}
\dot{x}&=&f_x(x,y,z)+\sqrt{\sigma_x}\thinspace g_x(x)\thinspace
\xi^x(t) + D(<x>-x),
\label{mean_eq_1}\\
\dot{y}&=&f_y(x,y,z)+\sqrt{\sigma_y}\thinspace g_y(y)\thinspace
\xi^y(t) + D(<y>-y),
\label{mean_eq_2}\\
\dot{z}&=&f_z(x,y,z)+\sqrt{\sigma_y}\thinspace g_z(y)\thinspace
\xi^z(t) + D(<z>-z), \label{mean_eq_3}
\end{eqnarray}
where $<x>$, $<y>$ and $<z>$ are average values on the spatial
lattice considered (ensemble averages in the thermodynamic limit)
and we set $f_x(x,y,z)=\lambda x(1-\nu x-\beta y-\alpha z)$,
$g_x(x)=x$, $f_y(x,y,z)=\lambda y (1-\nu y-\beta x-\alpha z)$,
$g_y(y)=y$, $f_z(x,y,z)=\lambda_z z[-1+\gamma(x+y)]$, $g_z(z)=z$. By
site averaging Eqs.~(\ref{mean_eq_1})-(\ref{mean_eq_3}), we obtain
\begin{equation}
<\dot{x}>~=~<f_x(x,y,z)>,~~<\dot{y}>~=~<f_y(x,y,z)>,~~<\dot{z}>~=~<f_z(x,y,z)>.
\label{mean site eq}
\end{equation}
By expanding the functions $f_x(x,y,z)$, $g_x(x)$, $f_y(x,y,z)$,
$g_y(y)$, $f_z(x,y,z)$, $g_z(z)$ around the $1^{st}$ order moments
$<x(t)>$, $<y(t)>$ and $<z(t)>$, we get an infinite set of
simultaneous ordinary differential equations for all the
moments~\cite{Kawai}. To truncate this set we apply a Gaussian
approximation, for which the cumulants above the $2^{nd}$ order
vanish. Therefore we obtain
\begin{eqnarray}
<\dot{x}>&=&\lambda<x>(1-\nu\thinspace<x>-\beta\thinspace<y>-\alpha\thinspace<z>)-\lambda(\nu\thinspace\mu_{200}+\beta
\thinspace\mu_{110}+\alpha\thinspace\mu_{101})\qquad\label{x_mean_eq}\\
<\dot{y}>&=&\lambda<y>(1-\nu\thinspace<y>-\beta\thinspace<x>-\alpha\thinspace<z>)-\lambda(\nu\thinspace\mu_{020}+\beta\thinspace
\mu_{110}+\alpha\thinspace\mu_{011})\qquad\label{y_mean_eq}\\
<\dot{z}>&=&\lambda_z<z>(-1+\gamma\thinspace<x>+\gamma\thinspace<y>)+\lambda_z\gamma(\mu_{101}+\mu_{011})\qquad\label{z_mean_eq}\\
\dot{\mu}_{200}&=&2\lambda(1-2\nu \thinspace<x>-\beta \thinspace<y>-\alpha \thinspace<z>)\mu_{200}\nonumber\\
&-& 2\lambda <x>(\beta \thinspace \mu_{110}+\alpha \thinspace \mu_{101})+2 \sigma_x (\mu_{200}+<x>^2)-2 D \mu_{200}\quad\label{mu200_eq}\\
\dot{\mu}_{020}&=&2\lambda(1-2\nu \thinspace <y>-\beta \thinspace <x>-\alpha \thinspace <z>)\mu_{020}\nonumber\\
&-& 2\lambda \thinspace <y>(\beta \thinspace \mu_{110}+\alpha \thinspace \mu_{011})+2 \sigma_y (\mu_{020}+<y>^2)-2 D \mu_{020}\quad\label{mu020_eq}\\
\dot{\mu}_{002}&=&2\lambda_z(-1+\gamma \thinspace <x>+\gamma \thinspace <y>)\mu_{002}\nonumber\\
&+& 2\lambda_z \gamma <z> (\mu_{101}+\mu_{011})
+2\sigma_z(\mu_{002}+<z>^2) - 2 D \mu_{002}\quad\label{mu002_eq}\\
\dot{\mu}_{110} &=& \lambda[2-2\nu\thinspace(<x>+<y>)-\beta\thinspace(<x>+<y>)-2\alpha\thinspace<z>]\mu_{110}\nonumber\\
&-&\lambda \beta(<x> \thinspace \mu_{020}+<y> \thinspace \mu_{200})
-\lambda\alpha(<x>\thinspace\mu_{011}+<y> \thinspace \mu_{101})
-2D\mu_{110}\quad\label{mu110_eq}\\
\dot{\mu}_{101} &=&
\lambda(1-2\nu\thinspace<x>-\beta\thinspace<y>-\alpha <z>)\mu_{101}
+\lambda_z(-1+\gamma \thinspace <x>+\gamma \thinspace <y>)\mu_{101}\nonumber\\
&-&\lambda <x> (\alpha \thinspace \mu_{002}+\beta \thinspace
\mu_{011}) +\lambda_z\gamma <z>(\mu_{110}+\mu_{200})-2D
\mu_{101}\quad\label{mu101_eq}\\
\dot{\mu}_{011} &=& \lambda(1-2\nu <y>-\beta <x>-\alpha
<z>)\mu_{011}
+\lambda_z(-1+\gamma \thinspace <x>+\gamma \thinspace <y>)\mu_{011}\nonumber\\
&-&\lambda <y> (\alpha \thinspace \mu_{002}+\beta \thinspace
\mu_{101}) +\lambda_z\gamma <z>(\mu_{110}+\mu_{020})-2D
\mu_{011}.\quad\label{mu011_eq}
\end{eqnarray}
where $\mu_{200}$, $\mu_{020}$, $\mu_{002}$, $\mu_{110}$,
$\mu_{101}$, $\mu_{011}$ are the $2^{nd}$ order central moments
defined on the lattice
\begin{eqnarray}
\mu_{200}(t)&=&<x^2>-<x>^2,~~~~~~~~\mu_{020}(t)\;=\; <y^2>-<y>^2,~~~~~~~~\mu_{002}(t)\;=\; <z^2>-<z>^2,\nonumber\\
\mu_{110}(t)&=&<xy>-<x><y>,~~~~\mu_{101}(t)\;=\;
<xz>-<x><z>,~~~~\mu_{011}(t)\; =\; <yz>-<y><z>.\qquad
\end{eqnarray}
%
 In order to get the dynamics of the three species we
analyze the time evolution of the $1^{st}$ and $2^{nd}$ order
moments according to Eqs.~(\ref{x_mean_eq})-(\ref{mu011_eq}). As
initial conditions we consider each species uniformly distributed on
the spatial domain, that is we set $<x(0)>$ = $<y(0)>$ = $0.1$,
$<z(0)>$ = $2.0$, $\mu_{200}(0)$ = $\mu_{020}(0)$ = $\mu_{002}(0)$ =
$\mu_{110}(0)$ = $\mu_{101}(0)$ = $\mu_{011}(0)$ = $0$. Therefore,
from Eqs.~(\ref{x_mean_eq})-(\ref{mu011_eq}) we get, in the
deterministic case, the stationary values for $<x>$, $<y>$ and $<z>$
\begin{equation}
<x>^{stat}~=~<y>^{stat}~=~\frac{1}{2\gamma}, ~~<z>^{stat}~=~
\frac{2\gamma-(\beta_{down}+\nu)}{2\alpha\gamma}. \label{z_stat_bis}
\end{equation}
Using for the parameters the same values of the single site
analysis, we obtain $<x>^{stat}~=~<y>^{stat}~=~0.5$ and
$<z>^{stat}~=~1.5.$ We also fix the delay time at the same value
$\tau_d = 435$ used in the single site case. Finally, by numerical
integration of Eqs.~(\ref{x_mean_eq})-(\ref{mu011_eq}), setting
$D=10^{-1}$, we get the time series of the $1^{st}$ and $2^{nd}$
order moments for the following values of multiplicative noise
intensity $\sigma = 0$, $10^{-12}$, $10^{-6}$, $10^{-3}$. The
results are reported in
Figs.~\ref{moment_series_1},~\ref{moment_series_2}. Here we note
that, after a transient, the mean values of the two prey densities
(see panels a and d of
Figs.~\ref{moment_series_1},~\ref{moment_series_2}) oscillate around
the stationary values. The oscillations are connected with the
presence of two stable equilibrium points. For
$\beta=\beta_{down}<1$ the stable equilibrium is given by the
contemporary presence of the three species (coexistence regime).
Conversely, for $\beta=\beta_{up}>1$ the system goes towards a new
equilibrium point, with the predator tending to disappear (exclusion
regime). In the presence of a dynamical regime (the system switches
periodically from coexistence to exclusion), we observe the
appearance of correlated oscillations in the time series of
$<x(t)>$, $<y(t)>$ and $<z(t)>$. In particular, we note that
$<z(t)>$ is subject to oscillations occurring around a value smaller
than the stationary one ($<z>^{stat}=1.5$) and characterized by a
larger amplitude with respect to $<x(t)>$ and $<y(t)>$. This
behaviour is analogous to that observed in the case of single site
dynamics.
In the absence of noise (top of Fig.~\ref{moment_series_1}), the
time series of $<x(t)>$, $<y(t)>$ and $<z(t)>$ (panel a),
$\mu_{200}(t)$, $\mu_{020}(t)$, $\mu_{002}(t)$ (panel b) and
$\mu_{110}(t)$, $\mu_{101}(t)$, $\mu_{011}(t)$ (panel c) are
completely overlapped and each species maintains a homogeneous
distribution over the lattice, that is all the $2^{nd}$ order
moments remain equal to zero. For $\sigma=10^{-12}$ (bottom of
Fig.~\ref{moment_series_1}) no changes are observed in the behaviour
of the mean values (see panel d), and the variances of the three
species show correlated oscillations (panel e). In panel f,
$\mu_{110}$ oscillates taking on only negative values. This
indicates that the spatial distributions in the lattice will be
characterized by the presence of regions where species $x$ or
species $y$ prevails. The two preys will be distributed therefore in
non-overlapping spatial patterns. This picture is in agreement with
previous results obtained with a different model~\cite{Fiasconaro}.
Conversely, $\mu_{101}$ and $\mu_{011}$ are always zero (see panel f
of Fig.~\ref{moment_series_1}). This behaviour indicates that the
predator is uncorrelated with the density of each prey: the species
$z$ tends to occupy indifferently the sites where $x$ or $y$
prevails (see the time behaviour of $\mu_{002}$ in panel e of
Fig.~\ref{moment_series_1}), but is correlated with the total prey
density (a global increase of food availability improves the life
conditions of the predator). This explains why the variance of the
predator shows small oscillations. On the other hand, when exclusion
regime takes place, the two preys tend to occupy different sites,
"spreading out" in the spatial domain and causing an increase of
their variances (see panel e of Fig.~\ref{moment_series_1}) with a
stronger anticorrelation (see the behaviour of $\mu_{110}$ in panel
f of Fig.~\ref{moment_series_1}). Finally we note that the amplitude
of the oscillations both of the variances and covariances increases
as a function of the noise intensity: in particular they have the
same order of magnitude of $\sigma$ (see panels b, c, e, f, in
Figs.~\ref{moment_series_1},~\ref{moment_series_2}). In fact, for
higher levels of multiplicative noise ($\sigma = 10^{-6},10^{-3}$)
the amplitude of the oscillations increases and the periodical
anticorrelated behaviour between the two preys becomes more evident.
Conversely, no modifications appear in the time series of the mean
values as a function of the multiplicative noise intensity (see
panels a, d in Figs.~\ref{moment_series_1},~\ref{moment_series_2}).

Even if it is related to a very different mechanism, this behavior
is similar to the stochastic resonance effect produced in population
dynamics, when the interaction parameter is subjected to an
oscillating bistable potential in the presence of additive
noise~\cite{Valenti,Valenti1}. We note that in the absence of
external noise ($\sigma = 0$) both populations coexist and the
species densities oscillate in phase around their stationary
value~\cite{Valenti}. This occurs identically in each site of the
spatial lattice (single site dynamics). The behavior of the mean
values reproduces this situation. For $\sigma \neq 0$, in the single
site dynamics we observe anticorrelated oscillations of $x$ and $y$
(preys). By site averaging these noise-induced oscillations (see
Ref.~\cite{Valenti}) we recover the average behavior obtained in the
absence of noise. This spatial auto-averaging effect explains why
the $1^{st}$ order moment behavior is independent on the external
noise intensity, while the $2^{nd}$ order moments give information
on "spreading" and anticorrelation of the species densities in the
spatial domain.

\section{Coupled Map Lattice Model}
\label{sec:2} \vskip-0.2cm In this section we adopt a different
approach to analyze the dynamics of the three species on the square
lattice defined in Section~\ref{sec:1}. We consider the time
evolution of our system by using a coupled map lattice (CML)
model~\cite{Kaneko}. In this formalism both correlated and
anticorrelated spatial patterns of the three interacting species
have been found~\cite{Fiasconaro}. Here we calculate the moments by
using the CML model. By this approach, the dynamics of the spatial
distributions of the three species is given by the following
equations
\begin{eqnarray}
x_{i,j}^{(n+1)}&=&\lambda x_{i,j}^{(n)} (1-\nu
x_{i,j}^{(n)}-\beta^{(n)} y_{i,j}^{(n)}-\alpha\thinspace
z_{i,j}^{(n)})+\sqrt{\sigma_x} x_{i,j}^{(n)} \xi_{i,j}^{x{(n)}} +
D\sum_\rho (x_{\rho}^{(n)}-x_{i,j}^{(n)}),
\label{CLM-Lotka_1}\qquad\\
y_{i,j}^{(n+1)}&=&\lambda y_{i,j}^{(n)} (1-\nu
y_{i,j}^{(n)}-\beta^{(n)} x_{i,j}^{(n)}-\alpha\thinspace
z_{i,j}^{(n)})+\sqrt{\sigma_y} y_{i,j}^{(n)} \xi_{i,j}^{y{(n)}} +
D\sum_\rho (y_{\rho}^{(n)}-y_{i,j}^{(n)}), \label{CLM-Lotka_2}\qquad\\
z_{i,j}^{(n+1)}&=&\lambda_z z_{i,j}^{(n)} (-1+\gamma\thinspace
x_{i,j}^{(n)}+\gamma\thinspace y_{i,j}^{(n)})+\sqrt{\sigma_z}
z_{i,j}^{(n)} \xi_{i,j}^{z{(n)}} + D\sum_\rho
(z_{\rho}^{(n)}-z_{i,j}^{(n)}),\qquad \label{CLM-Lotka_3}
\end{eqnarray}
where $x^{(n)}_{i,j}$, $y^{(n)}_{i,j}$ and $z^{(n)}_{i,j}$ denote
respectively the densities of prey $x$, prey $y$ and predator $z$ in
the site $(i,j)$ at the time step $n$. According to the notation
used for the mean field approach, $\lambda$, $\lambda_z$, $\nu$,
$\beta$, $\alpha$, $\gamma$ and $D$ represent the same quantities
defined in Section~\ref{sec:1}. $\xi_{i,j}^{x(n)}$,
$\xi_{i,j}^{y(n)}$, $\xi_{i,j}^{z(n)}$ are independent Gaussian
white noise sources with zero mean and unit variance. The
interaction parameter $\beta^{(n)}$ corresponds to the value of
$\beta(t)$ taken at the time step $n$, according to
Eq.~(\ref{jump_rate}). Here $\sum_\rho$ indicates the sum over the
four nearest neighbours.

\subsection{Stationary states for the CML model}
\vskip-0.2cm Applying a procedure analogous to that used for
Eqs.~(\ref{Lotka_eq_1})-(\ref{Lotka_eq_3}), we consider
Eqs.~(\ref{CLM-Lotka_1})-(\ref{CLM-Lotka_3}) in the absence both of
noise sources and diffusion terms ($D=0$). In this conditions, for
$x_{i,j}^{(n+1)}=x_{i,j}^{(n)}$, $y_{i,j}^{(n+1)}=y_{i,j}^{(n)}$,
$z_{i,j}^{(n+1)}=z_{i,j}^{(n)}$, we obtain the stationary values of
the three species densities for the generic site
\begin{eqnarray}
x_{_{CML}}^{stat}&=&y_{_{CML}}^{stat}=\frac{1}{2\gamma}\left[\frac{\lambda_z+1}{\lambda_z}\right]\label{CML_x_stat}\\
z_{_{CML}}^{stat}&=&\frac{2\gamma\left[\frac{\lambda-1}{\lambda}\right]-
(\beta+\nu)\left[\frac{\lambda_z+1}{\lambda_z}\right]}{2\alpha\gamma},\label{CML_z_stat}
\end{eqnarray}
where the indices $i,j$ were suppressed. As in the approach based on
the use of differential equations, the stationary values of the two
prey densities are independent on the interaction parameter $\beta$,
which is responsible for the two different dynamical regimes,
coexistence or exclusion, and affects the dynamics of the whole
system through its action on the stationary value
$z_{_{CML}}^{stat}$. The existence condition for the predator
\begin{equation}
z_{_{CML}}^{stat}=\frac{2\gamma\left[\frac{\lambda-1}{\lambda}\right]-
(\beta+\nu)\left[\frac{\lambda_z+1}{\lambda_z}\right]}{2\alpha\gamma}>0
\label{coex_CML_1}
\end{equation}
allows to get the following inequality for the interaction parameter
$\beta$
\begin{equation}
\beta<2\gamma
\frac{\left[\frac{\lambda-1}{\lambda}\right]}{\left[\frac{\lambda_z+1}{\lambda_z}\right]}-\nu.\label{coex_CML_2}
\end{equation}
The inequality (\ref{coex_CML_2}) indicates that, according to the
analysis performed in Section~\ref{single site dynamics}, the CML
model is characterized by two stationary states that become stable
or unstable depending on the values of the parameters. Comparing the
inequalities~(\ref{coex_2}) and~(\ref{coex_CML_2}), we note that in
the CML model the coexistence condition and the regulatory role,
played by the predator on the dynamics of the two preys, depend also
on the scale factors $\lambda$ and $\lambda_z$.

\subsection{Time series in the CML model}
\begin{figure}[htbp]
\begin{center}
\includegraphics[width=15cm]{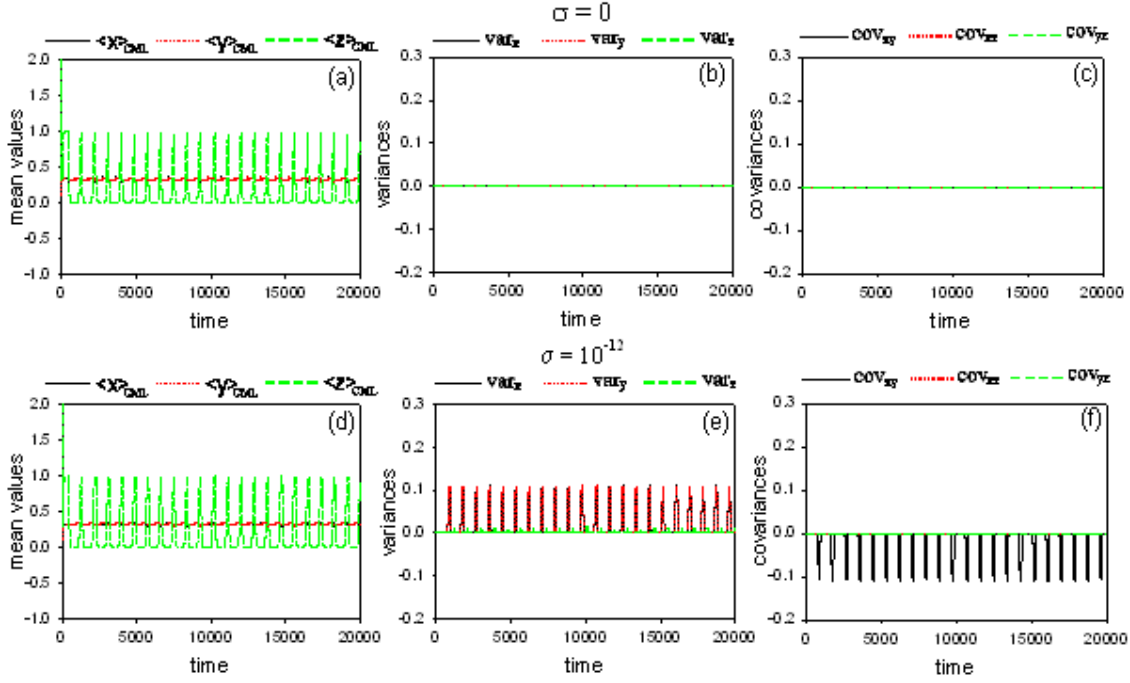}
\end{center}
\vskip-0.3cm \caption{\small In panels (a), (b) and (c) we show,
respectively, the mean values, $<x>^{(n)}$, $<y>^{(n)}$,
$<z>^{(n)}$, the variances, $var^{(n)}_x$, $var^{(n)}_y$,
$var^{(n)}_z$, and the covariances, $cov^{(n)}_{xy}$,
$cov^{(n)}_{xy}$, $cov^{(n)}_{xy}$ for $\sigma=0$. The same
quantities are shown in panels (d), (e) and (f) for
$\sigma=10^{-12}$. The time series are obtained within the formalism
of the CML model (see
Eqs.~(\ref{CLM-Lotka_1}),~(\ref{CLM-Lotka_2}),~(\ref{CLM-Lotka_3})).
The diffusion coefficient is $D=10^{-4}$, and $\gamma=26.5$. The
initial values of the species concentrations are $x_{i,j}^{(0)} =
y_{i,j}^{(0)} = 0.1$, $z_{i,j}^{(0)} = 2.0$ for all the sites
$(i,j)$. The values of the other parameters are the same of
Fig.~\ref{moment_series_1}: $\lambda=3$, $\lambda_z=0.06$, $\nu=1$,
$\alpha=0.02$.\bigskip} \label{CML_series_1} \vskip-0.4cm
\end{figure}

\begin{figure}[htbp]
\begin{center}
\includegraphics[width=15cm]{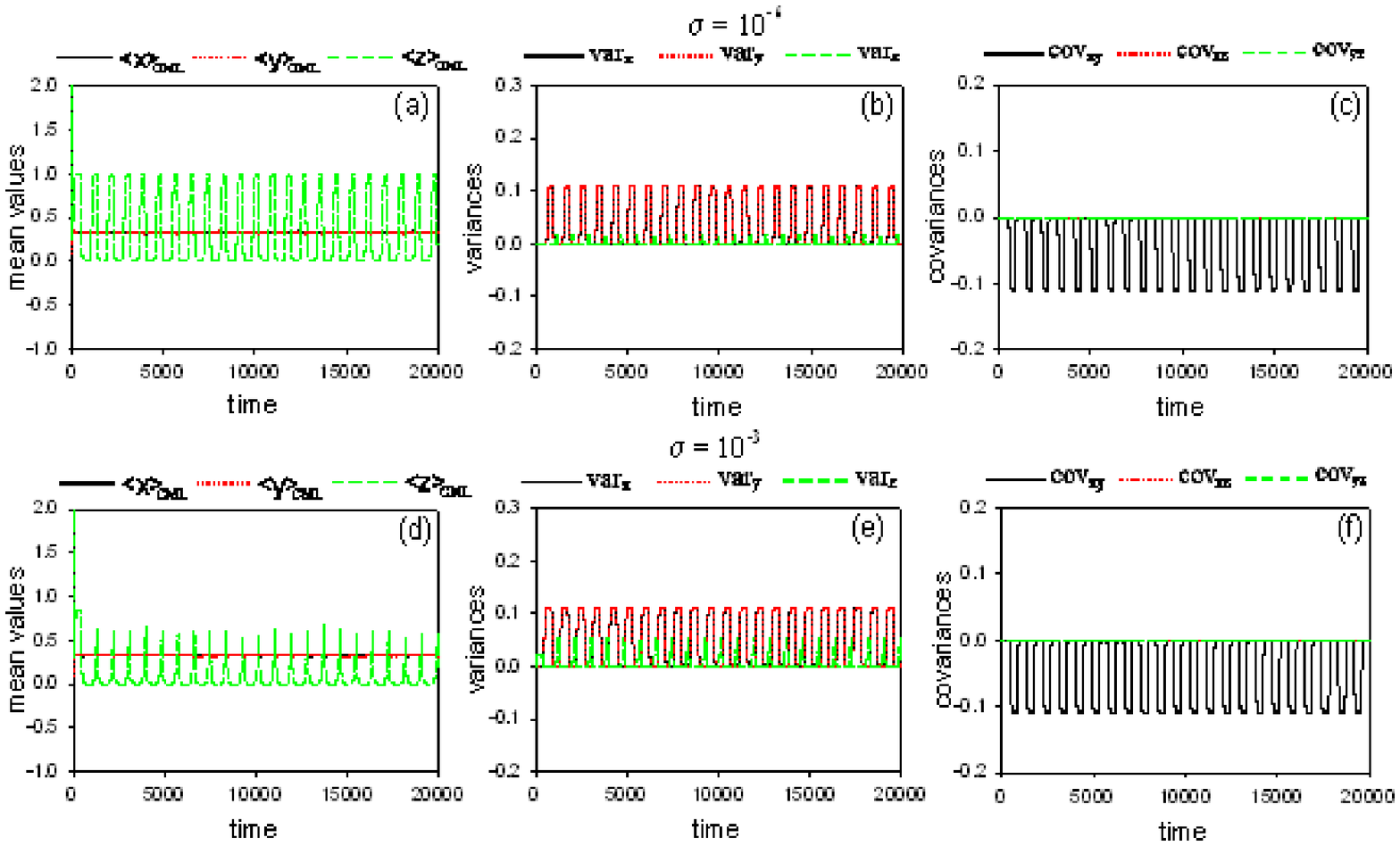}
\end{center}
\vskip-0.3cm \caption{ \small In panels (a), (b) and (c) we show,
respectively, the mean values, $<x>^{(n)}$, $<y>^{(n)}$,
$<z>^{(n)}$, the variances, $var^{(n)}_x$, $var^{(n)}_y$,
$var^{(n)}_z$, and the covariances, $cov^{(n)}_{xy}$,
$cov^{(n)}_{xy}$, $cov^{(n)}_{xy}$ for $\sigma=10^{-6}$. The same
quantities are shown in panels (d), (e) and (f) for
$\sigma=10^{-3}$. The time series are obtained within the formalism
of the CML model (see
Eqs.~(\ref{CLM-Lotka_1}),~(\ref{CLM-Lotka_2}),~(\ref{CLM-Lotka_3})).
The values of the other parameters and the initial conditions are
the same of Fig.~\ref{CML_series_1}.\bigskip}
 \label{CML_series_2}
\vskip-0.4cm
\end{figure}

In view of a comparison between mean field approach and CML model,
we define the $1^{st}$ and $2^{nd}$ order moments on the discrete
lattice, at the time step $n$. The mean values, $<x>^{(n)}$,
$<y>^{(n)}$, $<z>^{(n)}$, given by
\begin{equation}
<u>^{(n)}~=~\frac{\sum_{i,j}u^{(n)}_{i,j}}{N} \qquad\qquad\qquad (u
= x, y, z)\label{CMLmeans}
\end{equation}
represent the $1^{st}$ order moments. The variances $var_x^{(n)}$,
$var_y^{(n)}$, $var^{(n)}_{z}$ defined as
\begin{equation}
var_u^{(n)} =\frac{\sum_{i,j} (u^{(n)}_{i,j}-<u>^{(n)})^2}{N} \qquad
(u = x, y, z), \label{CMLvariances}
\end{equation}
and the covariances
\begin{eqnarray}
cov_{uw}^{(n)}=\frac{\sum_{i,j}(u^{(n)}_{i,j}-<u>^{(n)})
(w^{(n)}_{i,j}-<w>^{(n)})}{N} \enspace (u, w = x, y, z, \enspace
u\neq w) \label{CMLcovariances}
\end{eqnarray}
are the $2^{nd}$ order central moments. In order to get, for the
species densities, stationary values close to those obtained in the
mean field approach (see Eqs.~(\ref{z_stat_bis})), we choose for all
the parameters, except $\gamma$, the same values of
Section~\ref{sec:1}. Therefore, setting $\lambda=3$,
$\lambda_z=0.06$, $\nu=1$, $\alpha=0.02$, $\gamma=26.5$, from
Eqs.~(\ref{CML_x_stat}),~(\ref{CML_z_stat}), we calculate the
stationary values for the densities of the two preys and predator in
the coexistence regime ($\beta=\beta_{down}=0.94$)
\begin{equation}
<x>^{stat}_{CML}~=~<y>^{stat}_{CML}~=~0.3~;~~<z>^{stat}_{CML}~=~1.0.
\label{CML_z_stat_bis}
\end{equation}
The CML model can be considered as a time discrete version of the
Lotka-Volterra system, with time step $\Delta t=1$. For the
numerical integration of Eqs.~(\ref{x_mean_eq})-(\ref{mu011_eq}) we
used $dt=10^{-3}$, which is a suitable value to obtain convergence
of the solution. Obviously, with these values of $\Delta t$ and
$dt$, the dynamics of the CML model results to be faster with
respect to that obtained within the moment formalism. In particular,
for $\beta=\beta_{up}>1$, using the same parameter values of the
mean field approach, the exclusion regime causes the species $z$ to
vanish in one time step ($\Delta t=1$). This implies that, when the
system is subject to the dynamical regime discussed in
Section~\ref{single site dynamics}, the predator disappears. This
behaviour disagrees with the results found by using the moment
equations (see Section~\ref{spatially_extended}). In order to remove
this discrepancy between CML model and mean field approach, in the
discrete time equations we use a much smaller value for the
diffusion constant, namely $D=10^{-4}$. By this way, we obtain a
slowdown of the diffusion dynamics and, as a consequence, the
survivance of the predator in the coexistence/exclusion dynamical
regime. In order to get the time behaviour of the $1^{st}$ and
$2^{nd}$ order moments within the scheme of the CML model, we
consider a square lattice with $N = 100\times 100$, using for
$\beta(t)$ the time behaviour given in Fig.~\ref{beta}. Afterwards,
at each time step $n$ we calculate, from
Eqs.~(\ref{CLM-Lotka_1}),~(\ref{CLM-Lotka_2}),~(\ref{CLM-Lotka_3}),
the new values of $x_{i,j}^{(n)}$, $y_{i,j}^{(n)}$, $z_{i,j}^{(n)}$,
and the moments according to
Eqs.~(\ref{CMLmeans}),~(\ref{CMLvariances}),~(\ref{CMLcovariances}).
By iterating this procedure, we obtain the time series shown in
Figs.~\ref{CML_series_1},~\ref{CML_series_2}.
The $1^{st}$ and $2^{nd}$ order moments calculated within the
formalism of the CML model can be compared with the same quantities
obtained in the mean field approach (see
Figs.~\ref{moment_series_1},~\ref{moment_series_2}). We note that
the two set of time series are in a good qualitative agreement.
According to the results obtained in the formalism of the moment
equations, the mean values of the three species show time
oscillations, whose amplitude is larger for the predator (panels a,
d of Figs.~\ref{CML_series_1},~\ref{CML_series_2}). In the absence
of noise, the $2^{nd}$ order moments remain equal to zero (see panel
b of Fig.~\ref{CML_series_1}), recovering the conditions of
homogeneous distributions obtained for $\sigma=0$ in the mean field
approach (see panel b of Fig.~\ref{moment_series_1}). In the
presence of multiplicative noise, no modifications occur in the time
series of the $1^{st}$ order moments (see left side panels in
Figs.~\ref{CML_series_1},~\ref{CML_series_2}). However, for
$\sigma\neq 0$ a symmetry breaking is introduced, with non-vanishing
variances that are responsible for inhomogeneous distributions of
the three species. For higher levels of the noise intensity, the
amplitude of the oscillations remains constant in the time series of
$var_x$, $var_y$ and $var_z$ (see panels b, e of
Figs.~\ref{CML_series_1},~\ref{CML_series_2}). These results show
some difference with those obtained in the formalism of moment
equations, where higher noise intensities cause the oscillation
amplitudes of $\mu_{200}$, $\mu_{020}$ and $\mu_{002}$ to become
larger (see panels b, e of
Figs.~\ref{moment_series_1},~\ref{moment_series_2}). Finally, we
find that for $\sigma\neq 0$, temporal oscillations also appear in
the time series of $\mu_{110}$. This agrees with the results of the
mean field approach, revealing the presence of an anticorrelated
dynamics between the two preys. On the other hand, $\mu_{101}$ and
$\mu_{011}$ remain equal to zero also in the presence of
multiplicative noise. This behaviour, in agreement with that
obtained in the mean field formalism, indicates that the spatial
distribution of the predator is uncorrelated with those of each prey
considered separately, but depends on the total density of preys.
The comparison between the two approaches shows that the mean values
$<x>^{(n)}$, $<y>^{(n)}$, $<z>^{(n)}$ and those obtained within the
formalism of the moment equations oscillate around different values.
Moreover, the amplitudes of the oscillations in the $2^{nd}$ order
moments appear significantly larger in the CML model. This
discrepancies can be explained recalling that: i) in the two
approaches the stationary values are different (see
Eqs.~(\ref{z_stat_bis}) and
Eqs.~(\ref{CML_x_stat})-(\ref{CML_z_stat})); ii) in the mean field
formalism the interaction between sites is extended to the whole
spatial domain, conversely in the CML model the species interaction
is restricted to the nearest neighbors; iii) the dynamics of the CML
model is faster since an unitary time step ($\Delta t=1$) is taken,
instead of the time step $dt=10^{-3}$ used in the moment equations.

\section{Conclusions}
\label{concl} \vskip-0.2cm We report a study on the stochastic
dynamics of an ecosystem with three interacting species (two preys
and one predator), described by generalized Lotka-Volterra
equations. After considering the single site dynamics of the
ecosystem, we consider a spatially extended domain (two-dimensional
lattice) by introducing "long range" diffusive terms (diffusion
occurs among each site and all the other ones). The study is
performed by a mean field approach, in the formalism of the moment
equations. The system is affected by the presence of two noise
sources, namely a multiplicative white noise and a correlated
dichotomous noise. The role of the correlated dichotomous noise is
to control the dynamical regime of the ecosystem (see
Fig.~\ref{beta}), while the multiplicative noise is responsible for
the anticorrelated behavior of the species concentrations. The mean
field approach in Gaussian approximation enables us to obtain the
time series of the $1^{st}$ and $2^{nd}$ order moments. We compare
the results obtained within the mean field approach with the time
series calculated by a coupled map lattice (CML) model. The
agreement is quite good, even if some discrepancies are present, due
to the discrete nature of the CML model and the limited extension of
the diffusive interaction (nearest neighbors) among different sites
of the coupled map lattice. Our theoretical results could explain
the time evolution of populations in real ecosystems whose dynamics
is strictly dependent on random fluctuations, always present in
natural environment~\cite{Garcia,Caruso,Sprovieri}.

\section{Acknowledgments}
\label{aknow} \vskip-0.2cm Authors are thankful to Prof. Dr. Lutz
Schimansky-Geier that inspired this work by fundamental ideas and
useful discussions. Authors acknowledge the financial support by ESF
(European Science Foundation) STOCHDYN network and partially by
MIUR.

\end{document}